\documentclass[twocolumn,showpacs,%
  nofootinbib,aps,superscriptaddress,%
  eqsecnum,prd,notitlepage,showkeys,10pt]{revtex4-1}

\usepackage{amssymb}
\usepackage{amsmath}
\usepackage{graphicx}
\usepackage{dcolumn}
\usepackage{hyperref}
\usepackage{tabularx}
\usepackage{bera}% optional: just to have a nice mono-spaced font
\usepackage{listings}
\usepackage{xcolor}
\usepackage[pict2e]{struktex}

\colorlet{punct}{red!60!black}
\definecolor{background}{HTML}{EEEEEE}
\definecolor{delim}{RGB}{20,105,176}
\colorlet{numb}{black}

\lstdefinelanguage{json}{
    basicstyle=\normalfont\ttfamily,
    numbers=left,
    numberstyle=\scriptsize,
    stepnumber=1,
    numbersep=8pt,
    showstringspaces=false,
    breaklines=true,
    frame=lines,
    backgroundcolor=\color{background},
    literate=
     *{0}{{{\color{numb}0}}}{1}
      {1}{{{\color{numb}1}}}{1}
      {2}{{{\color{numb}2}}}{1}
      {3}{{{\color{numb}3}}}{1}
      {4}{{{\color{numb}4}}}{1}
      {5}{{{\color{numb}5}}}{1}
      {6}{{{\color{numb}6}}}{1}
      {7}{{{\color{numb}7}}}{1}
      {8}{{{\color{numb}8}}}{1}
      {9}{{{\color{numb}9}}}{1}
      {:}{{{\color{punct}{:}}}}{1}
      {,}{{{\color{punct}{,}}}}{1}
      {\{}{{{\color{delim}{\{}}}}{1}
      {\}}{{{\color{delim}{\}}}}}{1}
      {[}{{{\color{delim}{[}}}}{1}
      {]}{{{\color{delim}{]}}}}{1},
}

\begin{document}

\title{Right to Sign: Safeguarding data immutability in blockchain systems with cryptographic signatures over a broad range of available consensus finding scenarios}
\author{Ernst-Georg Schmid \href{mailto:ernst-georg.schmid@bayer.com}{ernst-georg.schmid@bayer.com}} 
\affiliation{Bayer Business Services GmbH}

\begin{abstract}
\textbf{Abstract.} The choice of the consensus method ultimately determines throughput, scalability, tamper resistance, and consistency of a blockchain system. However, across all the types of blockchain (private, semi-private, consortium, or public), there is no consensus method that uniformly addresses all these traits. Verifiable lottery algorithms (Proof of ...) increase tamper resistance but show weakness in throughput and scalability, while established methods like PAXOS and RAFT provide no additional protection against tampering. In this paper, we introduce \textit{Right to Sign} which aims to provide additional tamper resistance by cryptographic signatures over a broad range of available consensus finding methods.
\end{abstract}

\maketitle

\section{Introduction}

The publication of the seminal paper on Bitcoin \cite{nakamoto2008} has sparked an ongoing global interest in cryptocurrencies and the underlying blockchain technology. Recent developments like Ethereum \cite{ethereum} and Hyperledger \cite{hyperledger} aim for making blockchain technology availabe to a broader range of applications that require data immutability and tamper resistance, e.g. medical records, track \& trace in supply chains, or identity \& access management. But like other distributed database systems, blockchains suffer from the limited scalability of finding distributed consensus on who is allowed to write data \cite{vukolic}, which is imperative to keep data consistent \cite{bernstein, brewer}.

\section{Blockchain}

\subsection{Definition of blockchain}
We define a blockchain as a sequence of data containers called blocks $B_{0}, B_{1}, \ldots , B_{n}$. $B_{0}$ is called the genesis block. Each block can be identified by its hash, which is the hash value of some or all data in the block. Furthermore, a block contains at least a block number that satisfies the clock consistency condition, and the hash of the previous block.

\begin{figure}[h]
\includegraphics[width=\columnwidth]{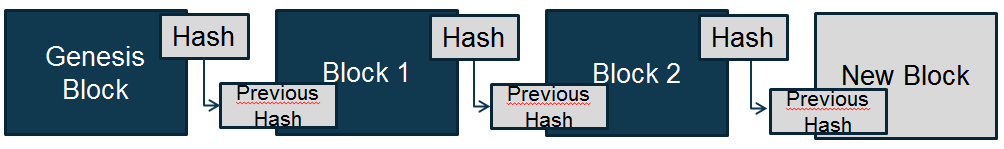}
\caption{\label{fig:blockchain_simple}A simple blockchain.}
\end{figure}

The blocks are then chained together by referencing their immediate predecessor hash, hence the name blockchain, as shown in Figure~\ref{fig:blockchain_simple}.

\subsection{Data immutability}
\label{subsec:data_immutability}
Data stored in a blockchain is considered to be immutable. To achieve immutability, a blockchain follows these basic principles:

\begin{enumerate}
	\item Blocks are only appended at the end of the chain.
	\item There has to be consensus on who is allowed to append the next block to the chain in order to avoid race conditions.
	\item Data already written is never changed. There is no UPDATE or DELETE.
	\item The hash of the previous block must be included in the current block.
\end{enumerate}

The integrity of every block, and hence the whole chain, can thus be verified by comparing its hash to the predecessor hash of the following block.

However, such a blockchain still can be changed if all block hashes from a changed block onwards can be rehashed with feasible effort. Therefore, additional safeguards are applied:

\begin{itemize}
	\item Copies of the blockchain are stored in multiple locations.
	\item The hash function to obtain a block hash is deliberately made computationally expensive.
	\item Additional attestation mechanisms, like digital signatures, can be used to further secure the block hashes.
\end{itemize}

\subsection{Tasks and nodes}
Following the principles given in \ref{subsec:data_immutability}, some basic tasks can be identified which must be fulfilled in a working blockchain system.

\begin{table}[h]
	\begin{tabularx}{\columnwidth}{X|X}
		\textbf{Task} & \textbf{Purpose}\\
		\hline
		Append & Create a new block and append it to the chain with consensus of the whole blockchain system\\
		\hline
		Attest & Attest that there truly was consensus of the whole blockchain system at the time the block was added\\
		\hline
	\end{tabularx}
	\caption{Basic tasks in a blockchain system}
	\label{tab:BasicTasksInABlockchainSystem}
\end{table}

Tasks are delegated to specialized nodes in order to achieve horizontal scalability and fault tolerance.

\section{Finding consensus}
\begin{figure}[h]
	\includegraphics[width=\columnwidth]{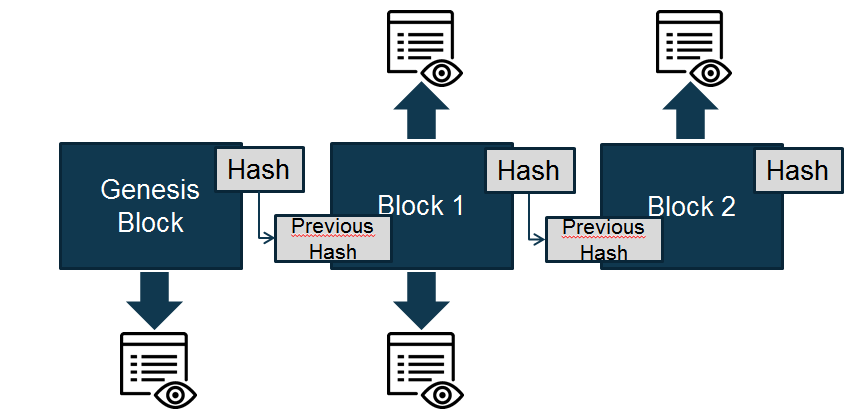}
	\caption{\label{fig:blockchain_reads}Concurrent reads in a Blockchain}
\end{figure}

\begin{figure}[h]
	\includegraphics[width=\columnwidth]{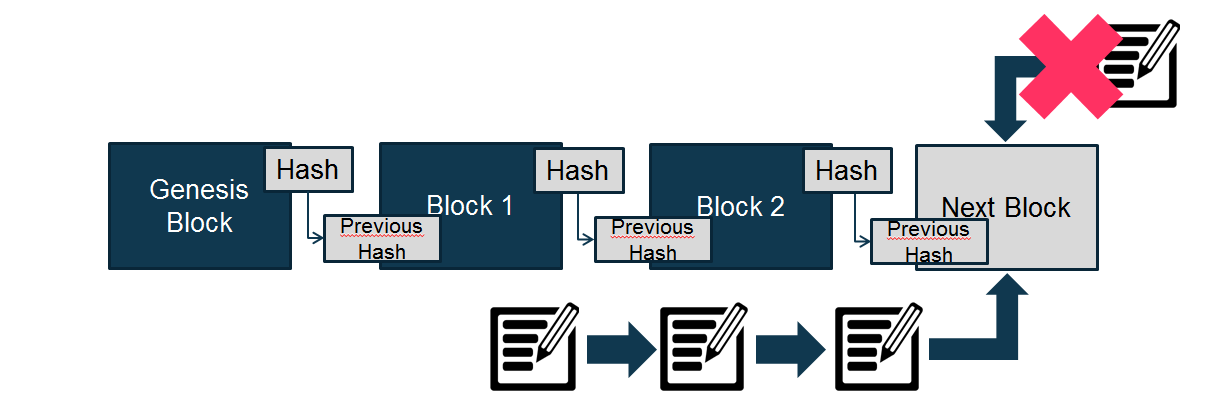}
	\caption{\label{fig:blockchain_writes}Serialized writes on a Blockchain}
\end{figure}

As already shown, there has to be consensus among all nodes on who is allowed to append the next block to the chain to avoid race conditions and potential data corruption. While concurrent reads can happen anywhere and anytime, as shown in Figure \ref{fig:blockchain_reads}, Figure \ref{fig:blockchain_writes} shows how concurrent writes must be serialized and must happen only at the end of the chain. 

As this is a common problem for all distributed database systems \cite{bernstein}, of which blockchain systems are one representative, diverse consensus finding algorithms for distributed systems have been, and are being, developed \cite{kumari}.

\subsection{Deterministic consensus}
In deterministic consensus algorithms, the leading node is chosen out of all participating nodes by a deterministic schedule. One example is the Round Robin method, where a node is selected by iterating over a set of $1 \ldots n$ nodes. For a Round-Robin schedule with $1 \ldots n$ nodes, $p=\frac{1}{n}$ if all nodes participate in every scheduling iteration.

A special case is the single node, where consensus is established simply by the lack of alternatives. For a single node, the probability $p$ to append the next block is obviously $p=1$.

\subsection{Voting consensus}
Voting consensus algorithms use elections to determine the leading node. This node is then the leader for the duration of an election term. Depending on the algorithm, the duration of an election term $t_{election}$ can be good for one action only, a fixed period of time, a random period of time or as long as this node does not fail. Voting consensus algorithms typically require comparatively extensive communication between the participating nodes, which means sending messages over some kind of network. RAFT \cite{raft} is a typical example. There is an excellent interactive explanation of RAFT that shows the amount of communication necessary to find consensus \cite{raft_explained}.

When sending messages over a network, there is usually no direct control over the latency of the network $t_{network}$ since this is subject to factors like overall network utilization, network failures, or malicious attacks. The probability of being elected as a leading node to append the next block is at least partially determined by two non-deterministic factors: 

$$p=P(t_{election}, t_{network})$$

\subsection{Proof of work consensus}
Proof of work was broadly introduced by \cite{nakamoto2008}. Unlike other consensus algorithms, it requires minimal network communication. The core idea is that any node can declare itself to be the leader and present a proof for this claim. This proof is then checked by every node and accepted or refused.

The original proof of work needs to iteratively find a nonce that, hashed together with the data in the block, produces a hash value which satisfies a global constraint, called difficulty $d$. This is computationally expensive, since there is no known way to find such a hash other than repeatedly 
hashing the data of the block and increment the nonce with each iteration. This hash is the block hash.

The validity of a block can be checked by all participating nodes by hashing the data and the nonce of the block once, and comparing the obtained hash value to $d$. This is an computationally cheap operation, compared to finding a block hash satisfying $d$ as described above.

So, proof of work uses a lottery mechanism where each node draws his own "`lucky numbers"' to find a valid block hash. If one is found, it is then communicated to all other nodes who in turn can verify the validity of the claim ex post, and then decide to either accept or refuse it.

The time period $T(r)$ for a node with hardware capable of performing $r$ hash operations per second to
find a valid block is distributed exponentially with the rate $\frac{r}{d}$
$$P\left\{T(r) \leq t\right\} = 1 - e^{-\frac{r}{d}\cdot t}$$
For $n$ nodes with hash rates $r_{1}, r_{2}, \ldots, r_{n}$ the period of time $T$ to find a valid block hash then equals the minimum value of random variables $T(r_{i})$ if the node publishes a valid block and
there is zero network latency. $T$
is then distributed exponentially also $$P\left\{T=T_{i}\right\} = \frac{r_{i}}{\sum\limits_{j=1}^n r_{j}}$$
This means any node with a share of compute power $s$ has the probability $p=s$ to find a valid block hash.

\section{The loreleiean beauty of proof of work consensus finding}
The beauty of proof of work is twofold. Unlike other consensus finding methods with changing leaders presented in this paper, proof of work minimizes network traffic during the election phase. Since every node can work on the Proof without communicating to others, network communication is only necessary for sending new blocks to the nodes and verification of the Proof, once a block shall be appended.
Also, proof of work also allows all nodes to verify ex post that an node really found the Proof by checking the block hash against the nonce and the difficulty $d$. So it is also self-attesting.

\begin{figure}[h]
\begin{lstlisting}[language=json,firstnumber=1]
{"block_number": 1,
"difficulty": 45323,
"nonce": 42,
"previous_block_hash": "2D711642B726B04401627CA9FBAC32F5C8530FB1903CC4DB02258717921A4881"}
\end{lstlisting}
\caption{\label{fig:pow_block_simple}Proof of work block header}
\end{figure}

Figure \ref{fig:pow_block_simple} shows a block header containing suffcient metadata to check for consensus and attestation of consensus by proof of work.

\subsection{Power consumption}

The main disadvantage of proof of work is, that is requires comparatively high compute resources to find the Proof. Current estimates for Bitcoin show, that it consumes 67 TW/h, or generates 32 kt of $CO_2$, as of May 2018 \cite{bitcoin_power}. One could say, that large scale proof of work is a ecological disaster.
Proof of work is also prone to race conditions if more than one node finds the Proof simultaneously. In case of Bitcoin, race conditions frequently lead to different copies of the blockchain which then have to merged back into one common one. Clients see already committed transactions disappear which have to be re-issued in order to not getting lost.

\subsection{Attacks against proof of work}
\label{subsec:PoW_attacks}
Proof of work is also susceptible to attacks where one or a coordinated group of participants accumulates at least 51\% of the total compute power. In this case, the attacker may deny attestation of other participant's transactions in favor of his own. Such attacks have been described in \cite{nakamoto2008} and are actually executed against cryptocurrencies with a comparatively low number of participants \cite{fiftyone}.

\section{Attestation in non proof of work based blockchain systems}
As shown before, proof of work combines consensus finding for the append and attest tasks in a blockchain system. If another consensus finding method is used, e.g. to mitigate the issues with proof of work, additional attestation must occur. A long-standing and proven method used for attestation are digital signatures \cite{digisig}.

\begin{figure}[h]
\includegraphics[width=\columnwidth]{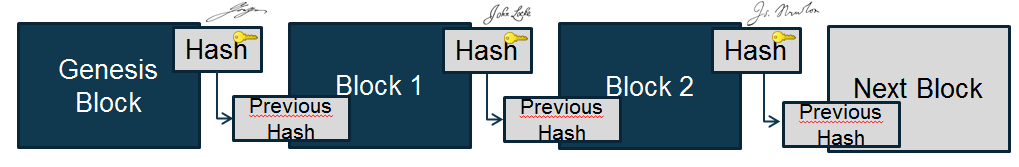}
\caption{\label{fig:blockchain_signed}A blockchain with attestation by digital signatures}
\end{figure}

\begin{figure}[h]
\begin{lstlisting}[language=json,firstnumber=1]
{"block_number": 1,
"certificate": "cmV0ZXRyZXdheGMgZmc0NDU1NjU2ZzY3ODY1NDN0cmVncmV6NTQ3aXRoam1oICB0ZXRlenJleg==",
"previous_block_hash": "2D711642B726B04401627CA9FBAC32F5C8530FB1903CC4DB02258717921A4881",
"signature": "A1FCE4363854FF888CFF4B8E7875D600C2682390412A8CF79B37D0B11148B0FA"}
\end{lstlisting}
\caption{\label{fig:sign_block_simple}Header of a digitally signed block}
\end{figure}

Figure \ref{fig:sign_block_simple} shows a simplified block header containing sufficient metadata to check for attestation of consensus by means of a digital signature.

\subsection{Digital signatures}
A digital signature is built by generating the hash value of some data, usually called a fingerprint $F$. The fingerprint is then encrypted with the secret key of the owner of this data which gives the signature $S$. 
$$hash(data) \rightarrow F_{owner}; encrypt(F_{owner}) \rightarrow S$$ 
This signature can be checked by any interested party by building $F$ again from the data received, decrypting the signature with the public key of the owner of this data and finally comparing both fingerprints. If they are not equal, the data has been changed after the signature was generated.
$$hash(data) \rightarrow F_{recipient}; decrypt(S) \rightarrow F_{owner};$$
$$check: F_{owner} = F_{recipient}$$

\subsection{Trust in digital signatures}
To forge a digital signature, an attacker would either have to find some alternative data that gives the same fingerprint, or generate a new signature. Because cryptographic hashing algorithms have very low collision probabilities and because of the asymmetric nature of public-key cryptography where the public key can only be used to encrypt and the private key only to decrypt or vice versa, both attack scenarios are very unlikely.
However, if the veracity of the owner's public key cannot be verified, any party checking the digital signature cannot be sure that the signature was truly built by the legitimate owner of the data. A well established means to provide such trust in identity without personally knowing the issuer of a digital signature are certificates.

A certificate is the public key of some entity (a person, a machine etc.) bundled together with other information about that entity, like name or address, that was itself digitally signed by a trusted third party, called a certificate authority (CA) to certify that it is genuine and bound to aforementioned entity.

The most widely used standard for certificates is defined by the ITU Telecommunication Standardization Sector of the International Telecommunication Union in recommendation X.509 \cite{X.509}, e.g. for SSL and TLS network transport encryption.

\section{Introducing Right to Sign}

Right to Sign separates the duties of finding consensus and attestation of consensus. Consensus is found either by an external consensus mechanism or proof of work. The attestation is always done by digital signatures.

To make this possible, we introduce a hybrid block header as shown in Figure \ref{fig:fusion_block_simple}.

\begin{figure}[h]
\begin{lstlisting}[language=json,firstnumber=1]
{"block_number": 1,
"difficulty": 45323,
"certificate": "cmV0ZXRyZXdheGMgZmc0NDU1NjU2ZzY3ODY1NDN0cmVncmV6NTQ3aXRoam1oICB0ZXRlenJleg==",
"previous_block_hash": "2D711642B726B04401627CA9FBAC32F5C8530FB1903CC4DB02258717921A4881",
"signature": "A1FCE4363854FF888CFF4B8E7875D600C2682390412A8CF79B37D0B11148B0FA"}
\end{lstlisting}
\caption{\label{fig:fusion_block_simple}A hybrid block header}
\end{figure}

It contains sufficient metadata to operate the blockchain in two modes of consensus finding, external or proof of work.
	
If global difficulty is set to zero, consensus must be found by an external method. The attestation is done by storing a certificate signed by a CA in the block and signing the blockhash with the private key of that certificate.

\begin{figure}[h]
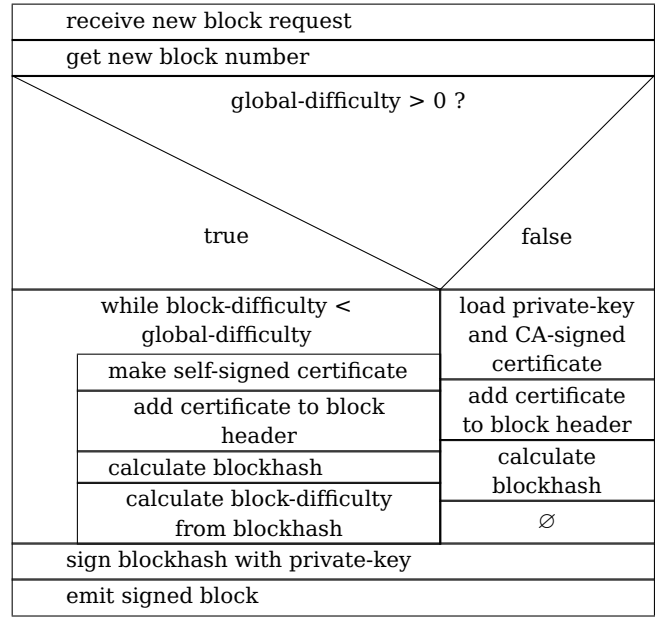

	\begin{struktogramm}(85,80)
		\assign{receive new block request}
		\assign{get new block number}
		\ifthenelse{2}{1}{global-difficulty > 0 ?}{true}{false}
		\while{while block-difficulty < global-difficulty}
		\assign{make self-signed certificate}
		\assign{add certificate to block header}
		\assign{calculate blockhash}
		\assign{calculate block-difficulty from blockhash}
		\whileend
		\change
		\assign{load private-key and CA-signed certificate}
		\assign{add certificate to block header}
		\assign{calculate blockhash}
		\ifend
		\assign{sign blockhash with private-key}
		\assign{emit signed block}
	\end{struktogramm}
	\caption{\label{fig:strukto}The Right to Sign algorithm}
\end{figure}

If global difficulty is greater than zero, consensus is found by repeated generation of a self-signed certificate, storing that certificate in the block, and making the blockhash until the block difficulty is equal or greater than the global difficulty.	Attestation is then done by signing the blockhash with the private key of the generated self-signed certificate.

Figure \ref{fig:strukto} describes Right to Sign in Nassi-Shneiderman notation. Concomitant proof-of-concept Python \cite{python} code was released \cite{R2S}.

Since attestation is now independent of the consensus finding method even mixing methods becomes possible. 
Right to Sign allows to build blockchain systems where a change of the consensus method does not force a change in the data structure of the blocks.

\subsection{Attacks against Right to Sign}
In proof of work mode, Right to Sign is vulnerable to 51\% attacks as described in \ref{subsec:PoW_attacks} like any other proof of work like method. For the case of Sybil attacks \cite{sybil_attacks}, proof of work inhibits them through the amount of compute resources needed. However, when using node bound certificates, an attacker could try a Sybil attack by adding rogue nodes to the system. But such nodes would have to bring their own certificates so that such an attack could be detected and stopped by rejecting new blocks signed with unknown certificates. If the list of allowed certificates is known to all regular nodes, any unknown certificate can be detected in $\mathcal{O}(1)$ time.

\section{Conclusion}

By decoupling the tasks of consensus and attestation, where attestation is always accomplished using digital signatures, Right to Sign allows to choose from a wide range of consensus methods, from fixed leader to proof of work. Together with the accompanying universal block header, this allows to build blockchain systems where a change of the consensus method does not force a change in the data structure of the blocks. We see this as a thought-provoking impulse towards decoupling data from protocol in future blockchain systems, aiming for better interoperability and scalability. 

\begin{acknowledgments}
A big thank you to Philip Gillissen and Hubertus Wortmann for their comments which greatly improved the manuscript.
\end{acknowledgments}

\onecolumngrid

\end{document}